\documentclass[aps,prl,twocolumn,groupedaddress]{revtex4}
\usepackage{graphicx}

\begin{document}

\title{\bf Sympathetic cooling of $^4$He$^+$ ions in a radiofrequency trap}
\author{B. Roth, U. Fr\"{o}hlich, and S. Schiller}
\affiliation{\it Institut f\"{u}r Experimentalphysik,
Heinrich--Heine--Universit\"{a}t D\"{u}sseldorf, 40225
D\"{u}sseldorf, Germany}

\begin{abstract}
\noindent We have generated Coulomb crystals of ultracold
$^4$He$^+$ ions in a linear radiofrequency trap, by sympathetic
cooling via laser--cooled $^9$Be$^+$. Stable crystals containing
up to $150$ localized He$^+$ ions at $\sim$20 mK were obtained.
Ensembles or single ultracold He$^+$ ions open up interesting
perspectives for performing precision tests of QED and
measurements of nuclear radii. The present work also indicates the
feasibility of cooling and crystallizing highly charged atomic ions
using $^9$Be$^+$ as coolant.
\end{abstract}
\maketitle
\noindent The two-body Coulomb system is one of the most
fundamental in physics, and has been central in the development of
quantum mechanics, relativistic quantum mechanics, and QED. In
nuclear physics, the study of these systems can provide an
alternative method for precise determination of nuclear sizes
\cite{Hatom}. Hydrogen--like systems studied include the hydrogen
atom and its isotopes, muonic hydrogen, the helium ions, muonium
and positronium. More recently, heavy (high--$Z$) hydrogen-like
ions have become available \cite{review_Heavy} and are being used
e.g. for exploring strong--field QED \cite{Verdu}, and for
measuring the electron mass \cite{Beier}. Among the low--$Z$
atomic systems, hydrogen has been the most extensively studied, in
particular by laser spectroscopy. This has resulted, among others,
in the most precise measurement of a fundamental constant, the
Rydberg constant \cite{Rydbergconstant}. While the helium ions
$^3$He$^+$ and $^4$He$^+$ are important systems because they are
complementary to the hydrogen atom, they have been much less
studied. Precision measurements of transition frequencies in
He$^+$ ions could provide (i) an independent (metrologically
significant) determination of the Rydberg constant, (ii) an
independent determination of the nuclear charge radii and the
isotope shift, assuming QED (Lamb shift) calculations are correct,
or (iii) a test of QED, using independent radius data (from
scattering measurements) as input.\\
\noindent Precise values of the He nuclear radii can test
theoretical nuclear methods and force models, which accurately
describe these special nuclei ($^3$He is the only stable
three-particle nucleus, $^4$He is lightest closed-shell nucleus).
The possibility of QED tests with He ions is attractive because
the QED corrections scale with high powers of $Z$ (some of the
theoretically unknown contributions scale as $Z^6$), and thus
their relative contribution to transition frequencies is larger
than in hydrogen. Also, the available independent $^4$He$^+$
radius measurements agree, in contrast to the situation in hydrogen.\\
\noindent On the experimental side, the hyperfine structure of
$^3$He$^+$ has been measured both in the electronic 1\,S ground
state and the 2\,S excited state \cite{hypfine3He}. The Lamb shift
of the 2\,S state in $^4$He$^+$ has been determined from the
spontaneous emission anisotropy with 170 kHz uncertainty and
compared to theoretical calculations
\cite{Wijngaarden2000,Jentschura2004}. The $^3$He/$^4$He squared
nuclear radius difference has been determined at the 0.4\% level
from the isotope shift in neutral
Helium using laser spectroscopy \cite{Shiner}.\\
\noindent A significant extension of these studies would be accessible via high--resolution laser
spectroscopy of helium ions, which has not been performed so far.
Measurements of the 1\,S - 2\,S and 2\,S - 3\,S two-photon
transition frequencies (at 61\,nm and 328\,nm, with linewidths
167\,Hz and 16\,MHz, respectively) have been proposed
\cite{Udem,Burrows}. For example, a 1\,S - 2\,S measurement with
reasonable experimental uncertainty $< 30$\,kHz would test the
nuclear and the recently improved theoretical QED contributions at
their current accuracy level \cite{accuracy}. A measurement of the isotope shift
with a similar accuracy would improve the value of the squared nuclear charge radius difference
by an order of magnitude.\\
One important aspect in these future precision experiments will be
the availability of trapped ultracold helium ions, in order to
minimize or eliminate the influence of Doppler broadening or
shifts, and to allow a precise study of systematic effects. The
experience with trapped ions for atomic clocks has shown the
success of this approach \cite{FSFSM2001}.\\
\noindent In this work we report on the first production of an
ultracold sample of $^4$He$^+$. While trapping of these ions is
straightforward using a Paul--type trap \cite{Major1968}, cooling
is more difficult. Direct laser cooling appears impractical at
present, since the generation of the required continuous--wave
deep--UV 30\,nm radiation is a challenging problem in itself. An
alternative and flexible method is sympathetic (interaction)
cooling, where ''sample'' particles of one species are cooled by
an ensemble of directly cooled (often by laser cooling) particles
of another species via their mutual interaction. This method was
first demonstrated for ions in Penning traps \cite{NIST1,NIST2},
and later in Paul traps \cite{Raizenetal92,Wakietal92,Bowe99}.
Under strong cooling, the mixed--species ensemble forms a Coulomb
crystal. In sympathetic crystallization of large ensembles a
minimum mass ratio range $m_{sc}$/$m_{lc}$ of 0.6 between
sympathetically cooled and laser--cooled ions has so far been
achieved \cite{Hornekaerthesis00}. Recent molecular dynamics
simulations showed that sympathetic cooling in an ion trap down to
a mass ratio of 0.3 should be possible
\cite{Harmonetal02,Schiller2003}. A two-ion sympathetically cooled
crystal exhibited a mass ratio of 0.38 \cite{Barrett2003}. In the
present work we have used the lightest atomic ion suitable for
practical laser cooling, $^9$Be$^+$, for which the mass ratio
to $^4$He$^+$ is 0.44.\\
\noindent We use a linear quadrupole trap to simultaneously store
Be$^+$ and He$^+$ ions. The trap is enclosed in a ultra-high
vacuum chamber kept below $1 \times 10^{-10}$ mbar and consists of
four cylindrical electrodes, each sectioned longitudinally into
three parts. The overall length of the electrodes is 10 cm, the
central region being 1.6 cm long. A necessary condition for stable
trapping of noninteracting ions is a Mathieu stability parameter,
$q = 2Q V_{RF}/m \Omega^2r^{2}_{0}$, $<$ 0.9. Here, $Q$ and $m$
are the charge and the mass of the trapped ions, $V_{RF}$ and
$\Omega$ are the amplitude and the frequency of the rf driving
field and $r_{0} = 4.3$\,mm is the distance from the trap
centerline to the electrodes. Trap operation at small $q$
parameters is favorable since rf--heating effects are less
pronounced. We operate our trap at an rf frequency of
$\Omega=2\pi$$\cdot$$14.2$\,MHz and an rf amplitude of 380\,V
where the stability parameter is $q \simeq$ 0.04 (0.1) for Be$^+$
(He$^+$) ions. For such small $q$ one can approximate the motion
of ions (in absence of interactions) by that in an effective
time-independent harmonic potential
\begin{equation}
U_{trap}(x,y,z) =\frac{m}{2}(\omega_r^2(x^2+y^2)+\omega_z^2 z^2)
\label{eq:pseudo}
\end{equation}
The $z$--axis is along the trap centerline. Oscillations
transverse to the $z$--axis occur with frequency $\omega_r =
(\omega_0^2 - \omega_z^2/2)^{1/2}$, with
$\omega_0=QV_{RF}/\sqrt{2}m\Omega r_0^2$ being the limiting value
for a very prolate trap. The longitudinal frequency $\omega_z = (2
\kappa Q V_{EC}/m)^{1/2}$ is obtained from a dc potential $V_{EC}$
applied to the 8 end sections (endcaps) of the electrodes, where
$\kappa \approx 3 \cdot 10^{-3}$/\,mm$^2$ is a
constant determined by the trap geometry.\\
\noindent For laser cooling of Be$^+$ ions we produce light
resonant with the ${^2}S_{1/2}(F=2)$ $\rightarrow$ ${^2}P_{3/2}$
transition at 313 nm. We employ doubly--resonant sum frequency
generation (SFG) of two solid--state laser sources, a resonantly
doubled Nd:YAG laser (532 nm), and a Ti:Sapphire laser at 760 nm,
in a bow--tie shaped ring cavity containing an LBO crystal
\cite{Schnitzler}. The Nd:YAG laser is the master laser and is
frequency stabilized to a hyperfine transition of molecular
iodine. By slaving the SFG cavity to the master laser and locking
the Ti:Sapphire to the cavity, the sum frequency wave is also
frequency stabilized. An AOM placed before the iodine
stabilization setup allows to shift the UV frequency within a
range of 340 MHz while maintaining absolute frequency stability.
When the iodine stabilization is switched off, the continuous
tuning range exceeds 16 GHz. UV output powers of up to 100 mW were
obtained. Since the Doppler cooling transition is not a closed
transition spontaneous emission to the metastable ground state
${^2}S_{1/2}(F=1)$ is possible. This population is removed using
repumping light red detuned by 1.250 GHz, produced by an
electrooptic modulator. To minimize the effect of radiation
pressure force on the crystal shape we usually use two
counterpropagating laser beams. The Be$^+$ fluorescence is
simultaneously recorded with a photomultiplier and a CCD camera.\\
\noindent Our basic procedure is as follows: first, we load the
trap with He$^+$ ions by leaking He gas into the vacuum chamber at
a pressure of $10^{-8}$ mbar and  ionizing it in situ by a 750 eV
electron beam crossing the trap center. The loading rate is
controlled by the partial pressure of neutral He gas and the
electron beam intensity. After allowing for the He pressure to
drop to the initial value, Be$^+$ ions are produced by ionizing
neutral atoms evaporated from a Be oven with the same electron
beam. During Be$^+$ loading the cooling and repumper lasers are
continuously scanned over a 5 GHz interval below resonance. If
necessary, mass--selective cleaning of the trap is applied to
eject heavy impurity particles prior to crystallization. To this
end, we add a static quadrupole potential $V_{DC}$ to the trap
until the $a$ parameters ($a = 4QV_{DC}/m \Omega^2r^{2}_{0}$) of
the unwanted species lie outside the Mathieu stability range.
Thus, stable trapping of particles
exceeding the mass of the atomic coolants is prevented.\\
One of the well--known signs for a phase transition from the fluid
ion plasma to an ordered crystal state is the appearance of a
sudden drop in the detected fluorescence attributed to the
reduction in the particle velocities and thus Doppler broadening.
Once this occurs, the cooling laser frequency is held constant at
a red detuning of approx. half the natural linewidth (60 MHz) from
the Be$^+$ resonance. Under this condition the crystals are
stable, with a particle loss half
time of $\sim$3 h.\\
\begin{figure}[t]
   \centering
   \includegraphics[height=7.5cm]{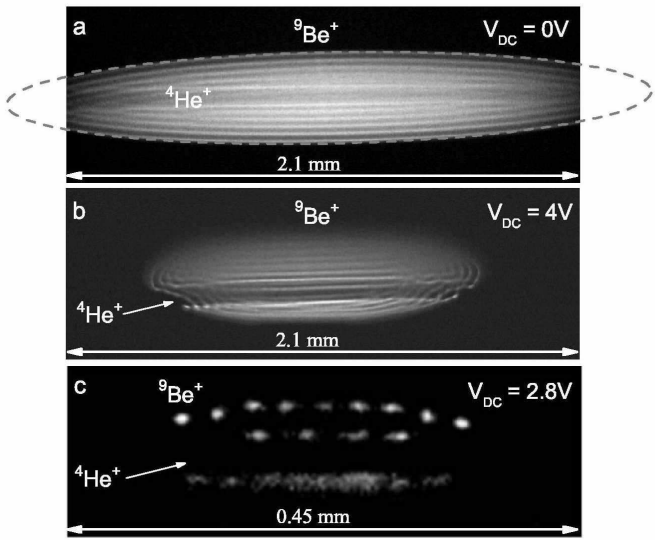}
   \caption{ \label{BeHe_Kristall_c} CCD images of two--component
Coulomb crystals (camera integration time: 2\,s). (a): spheroidal
crystal. The ellipse is a fit to the crystal boundary with a
semiaxes ratio $R/L=0.16$. (b,c): ellipsoidal crystal. The inner
dark cores contain approx. $150$, $30$, and $5$ sympathetically
crystallized He$^+$ ions, respectively. The trap axis $z$ is
horizontal. For (a,c) the cooling laser beam direction is to the
right, for (b) two counterpropagating beams were used. The
asymmetric ion distribution for (b,c) is attributed to static
stray potentials.}
\end{figure}
\noindent Fig.\ref{BeHe_Kristall_c} shows three prolate
two--component ion crystals containing Be$^+$ and He$^+$. The
crystals display the well-known shell structure and enclose an
inner dark region originating from the incorporated
sympathetically cooled particles having a lower mass--to--charge
ratio than that of the atomic coolants. Their location on--axis is
due to the stronger effective potential ($\sim$$Q^2/m$ for a
strongly prolate trap, $\omega_r \gg \omega_z$) experienced by
them as compared to the Be$^+$ ions, see Eq.(\ref{eq:pseudo}).
While Fig.\ref{BeHe_Kristall_c}\,a was taken with an axially
symmetric potential, in Fig.\ref{BeHe_Kristall_c}\,b,c a static
quadrupole potential $V_{DC}$ was added so as to make the region
occupied by the He$^+$ more visible. The added potential turns the
spheroidal crystal into an ellipsoid, squeezing the crystal in a
direction at 45$^\circ$ to the observation direction
\cite{Froehlich2004}. This results in a redistribution of the ions
such that fewer or no Be$^+$ ions remain in front and behind the
He$^+$ ions. The shapes of large Coulomb crystals containing a
small relative amount of He$^+$ ions agree well with the cold
fluid plasma model \cite{Turner,Froehlich2004}.\\
\noindent The number of crystallized particles is estimated by
performing molecular dynamics (MD) simulations and varying the
number of particles until the observed CCD image is reproduced.
While for single--species structures one may obtain the particle
number from the cold fluid model, which yields a particle density
${n_{0}} = \epsilon_0 V^2_{RF}/ m \Omega^2r_0^4$ ($\approx$$3.1
\times 10^4$/mm$^{3}$ for Be$^+$), together with measured
dimensions, here the MD approach is better suited. It is
applicable to multi--species structures, in addition produces
detailed structural information, and reduces uncertainties related
to the calibration of the imaging system magnification. For
example, MD simulations of the crystal in
Fig.\ref{BeHe_Kristall_c}\,a show that it contains approximately
$6.2\times10^3$ Be$^+$ ions and 150 He$^+$ ions. The radial
intershell distance is obtained as 29\,$\mu$m. This value agrees
well with the result calculated for infinite planar plasma
crystals, $1.48\,(3/4\pi n_{0})^{1/3} = 29.2$\,$\mu$m. According
to the simulations, the He$^+$ ions are arranged in a zig--zag
configuration along the trap axis for two--thirds of the crystal,
with a pitch spacing of $\approx$$40$\,$\mu$m. In the remaining
third (left end of the crystal) the He$^+$ ions form a linear
string, as evidenced by the smaller radial extension of the inner
dark core. The asymmetric distribution is caused by light pressure
forces on the Be$^+$ ions. Embedded strings were also observed in
a mixed crystal of Ca$^+$ and Mg$^+$ ions where both
ions were laser cooled \cite{Hornekaer00}.\\
\noindent An alternative way to roughly estimate the number of
sympathetically cooled He$^+$ ions is as follows. Crystallized
ions of any species are in force equilibrium at their locations
${\bf r}_i$. Thus, in a region occupied by a particular species,
the space charge electric field takes on the values $Q {\bf
E}({\bf r}_i) = \nabla U_{trap}({\bf r}_i)$. Consider now a closed
smooth surface that is chosen such that it closely approaches many
ion loci ${\bf r}_i$. We assume this relation approximately holds
on the entire surface. Applying Gauss' law, the total charge $Q_e$
enclosed by the surface is given by $Q_e=\epsilon_0 V_e \Delta
U_{trap}/Q$, where $V_e$ is the volume enclosed by the surface and
a harmonic trap has been assumed, $\Delta U_{trap}({\bf
r})=const.$ The Laplacian also determines the constant charge
density $\rho = \epsilon_0\Delta U_{trap}/Q = n_0 Q$ of an ion
species within the liquid charge model. Thus, the enclosed charge
is simply the "displaced charge", $Q_e=\rho V_e$. We now apply
this to the heavier particle species surrounding a cylindrical
core containing lighter ions. Taking as the surface the cylinder
bounded by the innermost Be$^+$--shell in
Fig.\,\ref{BeHe_Kristall_c}\,a (radius $r_{shell}\simeq 35\,\mu$m)
and extending to the ends of the dark core (length $\approx$2.2
mm), the estimate for the light species number is
$N_{He}=Q_e/Q_{He}=Q_{Be}n_{0,Be}V_e/Q_{He}\simeq 260$. This is in
reasonable agreement (within a factor of 2) with the MD
simulations, considering that $r_{shell}$ clearly varies over the
crystal length and the approximations made.\\
\noindent In order to identify the sympathetically cooled and
crystallized ions we have performed mass spectroscopy by
mass--selective excitation of transverse (secular) motion in the
trap. A plate electrode was placed between two trap rods and an ac
voltage applied (1\,V amplitude). A secular scan is taken by
varying the excitation frequency $\omega_{ext}/2\pi$ and recording
the Be$^+$ fluorescence with the photomultiplier. The secular
excitation pumps energy into the crystal; the resulting higher
Be$^+$ ion velocities imply a line broadening. This leads to less
fluorescence if the laser frequency lies near resonance, as is the
case when the detuning is optimized for maximum cooling. This
effect occurs for excitation of any species: since the species are
strongly coupled by Coulomb interaction, heating of one
species will lead to energy transfer to the others.\\
\begin{figure}[t]
\centering
   \includegraphics[height=12.5cm]{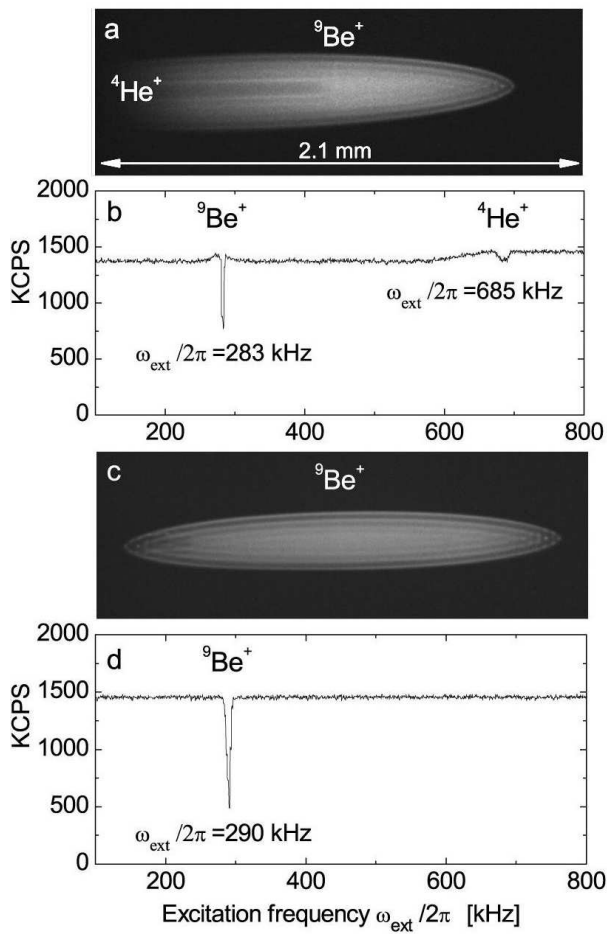}
   \caption{\label{BeHecrystal} (a): Prolate
two--component Coulomb crystal (cooling laser beam direction to
the right). (b): corresponding Be$^+$ fluorescence signal as a
function of secular excitation frequency $\omega_{ext}/2\pi$. (c):
ion crystal after removal of He$^+$ through repeated secular
excitation. (d): corresponding secular excitation spectrum. Time
between Fig. (a,b) and (c,d): 5 min. Frequency scans are toward
increasing frequency.}
\end{figure}
\noindent Fig.\ref{BeHecrystal} shows images for a large
two--component ion crystal before (a) and after (c) secular scans.
Initially, the secular scan shows the presence of both Be$^+$
(resonance at 283 kHz) and He$^+$ (685 kHz). The former value is
close to the calculated value of 271 kHz. Regarding the He$^+$
resonance, we notice both an increase in fluorescence before
resonance, and a substantial shift of the resonance peak from the
theoretical value (613 kHz). We attribute this spread and shift to
a modification of the secular frequency $\omega_r$, which is a
single-particle property, by the $z$--dependence of the Be$^+$ ion
distribution. The permanent increase of the fluorescence intensity
level before reaching the sharp He$^+$ secular excitation peak is
due to removal of He$^+$ ions from certain regions of the crystal
by the high excitation
amplitude.\\
\noindent After a few secular excitation cycles the
He$^+$ ions were nearly completely removed from the trap, Fig.
\ref{BeHecrystal}\,c, leaving behind a nearly pure Be$^+$ ion
crystal, as evidenced by the absence of secular resonance in
Fig.\,\ref{BeHecrystal}\,d. The small number of impurities
remaining in the left end of the crystal did not lead to a
fluorescence signal. Note that the visible crystal size has
decreased substantially because of Be$^+$ ion loss. Nevertheless,
the absolute fluorescence intensity of the final Be$^+$ crystal is
essentially equal to the level at the end of the first secular
scan. This may be explained by a lower ion temperature or reduced
micromotion. Moreover, a shift of the crystal with respect to the
trap occurred in the direction of the propagation of the laser
beam. We attribute this to stronger light pressure forces
experienced by the Be$^+$ ions, which is consistent with
the increased fluorescence level per ion.\\
\noindent A central question concerning sympathetically cooled
ions is their translational temperature; several observations can
shed light on the answer. A direct measurement on $^{24}$Mg$^+$
embedded in a laser--cooled $^{40}$Ca$^+$ crystal yielded an upper
limit of 45 mK, deduced from the $^{24}$Mg$^+$ laser excitation
linewidth \cite{Hornekaerthesis00}. In a two--ion crystal
sympathetic cooling to the Doppler--limit ($<$1 mK) has been shown
\cite{Barrett2003}. MD simulations for small particle numbers have
shown that sympathetic ions caged by laser-cooled ions are
essentially in thermal equilibrium with the latter at the Doppler
temperature \cite{Schiller2003}. Thus, an estimate of the
sympathetic temperature may be obtained from the temperature of
the laser--cooled ions. We deduce a direct upper limit for the
translational temperature of the Be$^+$ ions from the spectral
line shape of their fluorescence as the cooling laser is tuned
towards resonance and the ion ensemble crystallizes. Since the
temperature of the particles changes during the frequency scan, we
fit an appropriate Voigt profile to each point of the recorded
fluorescence curve to determine the Be$^+$ temperature. For small
crystals ($<$1000 particles), we find an upper limit for the
temperature at the end of the scan of 42\,mK. An indirect upper
limit is obtained by comparing the size of the ion spots with MD
simulations; here we find a tighter limit of $<$20 mK for the
Be$^+$ temperature. We therefore deduce, assuming thermal
equilibrium, that the He$^+$ temperature is $<$20 mK.\\
\noindent In summary, we have sympathetically cooled and
crystallized He$^+$ ions using laser--cooled Be$^+$ ions in a
linear Paul trap. Large Coulomb crystals of $\sim 6.2\times10^3$
Be$^+$ ions contained about $150$ He$^+$ ions, arranged in a
zig--zag structure centered on the trap axis. The mass ratio of
0.44 between sympathetically cooled and laser--cooled ions is the
lowest achieved so far for large ion ensembles in a Paul trap. We
estimate the temperature of the crystallized He$^+$ ions at
below 20 mK.\\
\noindent Translationally cold and immobilized He$^+$ ions are a
promising system for high precision spectroscopy and might lead to
more precise atomic and nuclear constants. The weak 1\,S - 2\,S
transition in He$^+$ could be detected with high signal to noise
ratio using a single adjacent Be$^+$ ion as a "quantum sensor",
rather than by direct detection of its fluorescence
\cite{Wineland}. Significant progress in this direction has been
achieved \cite{Blinov,Barrett2003}. Moreover, sympathetically
cooled He$^+$ opens up perspectives for studies of cold collisions
and cold chemistry, e.g. generation of $^3$HeH$^+$ molecules,
whose hyperfine
structure is of interest in fundamental physics \cite{Karsh}.\\
The $^4$He$^+$/$^9$Be$^+$ cooling and crystallization results also
have implications for the possibility of achieving this with
highly charged atomic ions (HCAI). Favorably, the $q$ parameters
are similar and the effective potential, Eq.\,(\ref{eq:pseudo}),
is steeper for the HCAIs than for the $^9$Be$^+$. Under
appropriate conditions, the laser--cooled $^9$Be$^+$ ions are
expected to cage the HCAIs,
leading to efficient cooling, as demonstrated with $^4$He$^+$.\\
\noindent We thank the Deutsche Forschungsgemeinschaft (DFG) and
the EU network HPRN-CT-2002-00290 for support. We are grateful to
D. Leibfried for helpful suggestions, H. Wenz for the MD
simulations, and S. Karshenboim for discussions.\\
Note added in proof: We have also produced two-species Coulomb
crystals containing $^3$He$^+$.


\end{document}